\documentclass[twoside, 10pt]{article}
\usepackage[greek,american]{babel}
\usepackage{amssymb}
\usepackage{amsfonts}
\begin{document}
\newtheorem{definition}{Definition}[section]
\newtheorem{remark}{Remark}[section]
\newtheorem{defintion}{Definition}[section]
\newtheorem{theorem}{Theorem}[section]
\newtheorem{proposition}{Proposition}[section]
\newtheorem{lemma}{Lemma}[section]
\newtheorem{help1}{Example}[section]
\title{A remark on the existence of breather solutions for the Discrete Nonlinear Schr\"odinger Equation in infinite lattices: The case of site dependent anharmonic parameter\footnote{\hspace*{0.5cm} AMS Subject Classification: 37L60, 35Q55, 47J30.}}
\author{Nikos I. Karachalios\\
{\it }\\
{\it Department of Mathematics,}\\
{\it University of the Aegean,}\\
{\it Karlovassi GR 83200, Samos, GREECE}
}
\date{}
\maketitle
\pagestyle{myheadings} 
\thispagestyle{plain}
\markboth{Nikos. I. Karachalios}{Breather solutions for DNLS, with site dependent anharmonic parameter}
\begin{abstract}
We discuss the existence of breather solutions for a Discrete Nonlinear Schr\"odinger equation in an infinite $N$-dimensional lattice, involving site dependent anharmonic parameter. We give a simple proof on the existence of (nontrivial) breather solutions based on variational approach, assuming that the sequence of anharmonic parameters is in an appropriate sequence space (decays with an appropriate rate). We also give a proof on the non-existence of (non-trivial) breather solutions, and discuss a possible physical interpretation of the restrictions, both in the existence and nonexistence case. 
\end{abstract}
\section{Introduction}
The one dimensional Discrete Nonlinear Schr\"odinger Equation (DNLS), 
\begin{eqnarray}
\label{DNLS} 
i\dot{\psi}_n+\epsilon(\psi_{n-1}-2\psi_n+\psi_{n+1})+\gamma |\psi_n|^2\psi_n=0,
\end{eqnarray}
represents an infinite ($n\in\mathbb{Z}$), or a finite ($|n|\leq K$), one-dimensional array of coupled anharmonic oscillators, coupled to their nearest neighbors with a coupling strength $\epsilon$. Here $\psi_n(t)$ stands for the complex mode amplitude
of the oscillator at lattice site $n$, and $\gamma$ denotes an anharmonic parameter. Setting $\epsilon=1/(\Delta x)^2$,
reminds that the model includes a finite spacing between molecules, and the formal continuum limit, the NLS partial differential equation, is obtained by taking $\Delta x\rightarrow 0$. The DNLS equation is one of the most inportant inherently discrete models, having a crucial role in modelling of a great variety of phenomena, ranging from solid state and condensed matter physics to biology, \cite{Aubry,Eil,FlachWillis,HennigTsironis,Yuri}. Depending on the size of the lattice, we have to deal with an infinite or finite system of ordinary differential equations, respectively.
 
The gauge invariance of the nonlinearity, allows for the support of special solutions of (\ref{DNLS}) of the form
$\psi_n(t)=\phi_n\exp (i\omega t)$, $\omega>0$. These solutions are called {\em breather solutions}, due to their periodic time behavior. Inserting the ansatz of a breather solution into (\ref{DNLS}), it follows that $\phi_n$ satisfies  the nonlinear system of algebraic equations
\begin{eqnarray}
\label{breather}
-\epsilon(\phi_{n-1}-2\phi_n+\phi_{n+1})+\omega\phi_n=\gamma |\psi_n|^2\psi_n.
\end{eqnarray}
The problem of existence and stability properties of breather solutions of coupled oscillators, has been developed as a fascinating sublect of research, from the derivation of the stationary DNLS equation \cite{Holstein}, the derivation of stationary solutions for the (coupled) DNLS,  by numerical continuation from the so-called anticontinuum limit (the case $\epsilon\rightarrow 0$) \cite{Eil2}, to the ingenious construction of localized time-periodic or quasiperiodic  solutions of general discrete systems, starting from periodic solutions of the corresponding anticontinuum limit equations \cite{Aubry, RSMackayAubry}. We refer to \cite{Eil,Kevrekidis} for a review of the existing results and the history of the problem as for a long list of references.

Motivated by \cite[Section 3]{Eil} and \cite{bang}, in this work we consider higher dimensional generalizations of DNLS, involving an arbitrary power law nonlinearity,  
and site dependence of the anharmonic parameter $\gamma$. For this particular case of nonlinearity, we also refer to \cite{Mol1,Mol2,Mol3}. For instance, we seek for breather solutions of the DNLS equation in infinite higher dimensional lattices ($n=(n_1,n_2,\ldots,n_N)\in\mathbb{Z}^N$),
\begin{eqnarray}
\label{DNLSh}
i\dot{\psi}_n+(\mathbf{A}\psi)_n+ \gamma_n|\psi_n|^{2\sigma}\psi_n=0, 
\end{eqnarray}
where 
\begin{eqnarray}
\label{DiscLap}
(\mathbf{A}\psi)_{n\in\mathbb{Z}^N}&=&\psi_{(n_{1}-1,n_2,\ldots ,n_N)}+\psi_{(n_1,n_{2}-1,\ldots ,n_N)}+\cdots+
\psi_{(n_1,n_{2},\cdots ,n_N-1)}\nonumber\\
&&-2N\psi_{(n_{1},n_2,\ldots ,n_N)}
+\psi_{(n_{1}+1,n_2,\ldots ,n_N)}\nonumber\\
&&+\psi_{(n_1,n_{2}+1,\ldots ,n_N)}+\cdots+
\psi_{(n_1,n_{2},\cdots ,n_N+1)},
\end{eqnarray}
In this case, equation (\ref{DNLSh}), could be viewed as the discretization of the NLS partial differential equation
\begin{eqnarray}
\label{NLSh}
i\psi_t+\Delta\psi+\gamma(x)|\psi|^{2\sigma}\psi=0,\;\;x\in\mathbb{R}^N.
\end{eqnarray}

As in the one dimensional case, it can be easily seen that any breather solution $\psi_n(t)=\phi_n\exp (i\omega t)$, 
of the DNLS equation (\ref{DNLSh}), satisfies the infinite nonlinear system of agebraic equations
\begin{eqnarray}
\label{swe}
-(\mathbf{A}\phi)_{n}+\omega\phi_n=\gamma_n|\phi_n|^{2\sigma}\phi_n,\;\;n\in\mathbb{Z}^N.
\end{eqnarray}
Based on a variational approach, which makes use of the famous  Mountain Pass Theorem (MPT),  we give a simple proof on the existence of (nontrivial) breather solutions for (\ref{DNLSh}), by showing that the energy functional associated to (\ref{swe}), has a critical point of ``mountain pass type''. 
Our main assumption is that $\gamma_n$ decays in an appropriate rate, in the sense that $\gamma_n$ is in an appropriate sequence space. This restriction enables us to use a compact inclusion between ordinary sequence spaces and {\em weighted} sequence spaces, in order to justify one of the crucial steps needed for for the application of MPT, namely the Palais-Smale condition. This is an important difference with the case of constant anharmonic parameter as the analysis of our recent work \cite{AN} shows: The latter is associated with lack of compactness, and restricted our study for a finite dimensional problem (in 1-D lattice, assuming Dirichlet boundary conditions). The application of MPT to (\ref{swe}) gives rise to some restrictions, which possibly have some physical interpretation, if viewed as local estimates for some ``energy quantities'' associated with the breather solution. 

On the other hand, it is shown that nontrivial solutions of (\ref{swe}) do not exist, in a sufficiently small ball of the energy space. The proof is based on a fixed point argument used also in \cite{AN}. This result could have the implementation, that we should not expect the existence of breather solutions, if the energy of the excitations of the lattice is sufficiently small.

If the estimates derived by the application of the MPT, do not appear just as a technical step for the proof,  they could be combined with that of the non-existence result, to derive a ``dispersion relation'' of nonlinearity exponent $\sigma$ vs the frequency $\omega$ of the breather solution, providing indication on the behavior of the associated energy quantities.
For a detailed discussion on the breather problem in higher dimensional lattices and the dependence of the frequency $\omega$ on the conserved quantities of DNLS, we refer to \cite[Section 6]{Eil}. 
%%%%%%%%%%%%%%%%%%%%%%%%%%%%%%%%%%%%
%%%%%%%%%%%%%%%%%%%%%%%%%%%%%%%%%%%%
%%%%%%%%%%%%%%%%%%%%%%%%%%%%%%%%%%%%
\section{Preliminaries}
In this introductory section, we describe the functional setting needed for the treatment of the infinite nonlinear system (\ref{swe}). We also introduce some weighted sequence spaces, and we prove a compact inclusion between the ordinary sequence spaces and their weighted counterparts.

For some positive integer $N$, we consider the complex sequence spaces 
\begin{equation}
%\begin{eqnarray}
\label{ususeqs}
\ell^p=\left\{
\begin{array}{ll}
&\phi=\{\phi_n\}_{n\in\mathbb{Z}^{N}},\;n=(n_1,n_2,\ldots,n_N)\in\mathbb{Z}^N,\;\;\phi_n\in\mathbb{C},\\
&||\phi||_{\ell^p}=\left(\sum_{n\in\mathbb{Z}^N}|\phi_n|^p\right)^{\frac{1}{p}}<\infty
\end{array}
\right\}.
\end{equation}
Between $\ell^p$ spaces the following elementary embedding relation  \cite{ree79} holds,
\begin{eqnarray}
\label{lp1}
\ell^q\subset\ell^p,\;\;\;\; ||\phi||_{\ell^p}\leq ||\phi||_{\ell^q}\,\;\; 1\leq q\leq p\leq\infty.
\end{eqnarray}
Note that the contrary holds for the $L^p(\Omega)$-spaces if $\Omega\subset\mathbb{R}^N$ has finite measure. For $p=2$, we get the usual Hilbert space of square-summable sequences, which becomes a real Hilbert space if endowed with the real scalar product
\begin{eqnarray}
\label{lp2}
(\phi,\psi)_{\ell^2}=\mathrm{Re}\sum_{{n\in\mathbb{Z}^N}}\phi_n\overline{\psi_n},\;\;\phi,\,\psi\in\ell^2.
\end{eqnarray}

For a sequence {\em of positive real numbers}\ \ $\delta=\{\delta_n\}_{n\in\mathbb{Z}^{N}}$,  we define the weighted sequence spaces $\ell^2_{\delta}$
\begin{equation}
\label{ususeqsw}
\ell^p_{\delta}=\left\{
\begin{array}{ll}
&\phi=\{\phi_n\}_{n\in\mathbb{Z}^{N}},\;n=(n_1,n_2,\ldots,n_N)\in\mathbb{Z}^N,\;\;\phi_n\in\mathbb{C},\\
&||\phi||_{\ell^p_{\delta}}=\left(\sum_{n\in\mathbb{Z}^N}\delta_n|\phi_n|^p\right)^{\frac{1}{p}}<\infty
\end{array}
\right\}.
\end{equation}
For the case $p=2$, it is not hard to see that $\ell^2_{\delta}$ is a Hilbert space, with scalar product
\begin{eqnarray}
\label{weightscal}
(\phi,\psi)_{\ell^2_{\delta}}=\mathrm{Re}\sum_{n\in\mathbb{Z}^N}\delta_n\phi_n\overline{\psi_n},\;\;\phi,\,\psi\in\ell^2_\delta.
\end{eqnarray}
For a certain class of weight $\delta$, we have the following lemma which shall play a crucial role in our analysis.
\begin{lemma}
\label{compactness}
We assume that the positive sequence of real numbers $\delta\in\ell^{\rho}$, $\rho=\frac{q-1}{q-2}$ for some $q>2$. Then
$\ell^2\hookrightarrow\ell^2_{\delta}$ with compact inclusion.
\end{lemma}
{\bf Proof:} We use the ideas of \cite[Lemma 2.3, pg. 79]{KJB} and (\ref{lp1}).  We consider a bounded sequence $\phi_k\in\ell^2$ and we denote by $(\phi_k)_n$ the $n$-th coordinate of this sequence. It suffices to show that the sequence $\phi_k$ is a Cauchy sequence in $\ell^2_{\delta}$. For some $q>2$ we consider its H\"older conjugate through the relation $p^{-1}+q^{-1}=1$. Then for all positive integers $k,l$, we have
\begin{eqnarray}
\label{lem1}
||\phi_k-\phi_l||^2_{\ell^2_{\delta}}&=&\sum_{n\in\mathbb{Z}^{N}}\delta_n|(\phi_k)_n-(\phi_l)_n|^2\\
&\leq& 
\left(\sum_{n\in\mathbb{Z}^{N}}\delta_n|(\phi_k)_n-(\phi_l)_n|^p\right)^{\frac{1}{p}}
\left(\sum_{n\in\mathbb{Z}^{N}}\delta_n|(\phi_k)_n-(\phi_l)_n|^q\right)^{\frac{1}{q}}.\nonumber
\end{eqnarray}
Since $\phi_k$ is a bounded sequence in $\ell^2$, it follows from (\ref{lp1}) that $\phi_k$ is bounded in $\ell^q$. 
Then from (\ref{lem1}) we have that there exists a positive constant $c$, such that
\begin{eqnarray}
\label{lem2}
||\phi_k-\phi_l||^2_{\ell^2_{\delta}}\leq c \left(\sum_{n\in\mathbb{Z}^{N}}\delta_n|(\phi_k)_n-(\phi_l)_n|^p\right)^{\frac{1}{p}}.
\end{eqnarray}
Since $\delta\in\ell^{\rho}$, it holds that for any $\epsilon_1>0$, there exists $K_0(\epsilon_1)$ such that for all $K>K_0(\epsilon_1)$ 
$$\sum_{|n|> K}|\delta_n|^{\rho}<\epsilon_1.$$
Thus, using  the boundedness of $\phi_k$ in $\ell^q$ once again, we have
\begin{eqnarray}
\label{lem5}
\sum_{|n|> K}\delta_n|(\phi_k)_n-(\phi_l)_n|^p&\leq& \left(\sum_{|n|> K}|\delta_n|^{\rho}\right)^{\frac{1}{\rho}}
\left(\sum_{|n|> K}|(\phi_k)_n-(\phi_l)_n|^q\right)^{\frac{p}{q}}\nonumber\\
&<& c\epsilon_1^{\frac{1}{\rho}}.
\end{eqnarray}
On the other hand, since 
the sequence $\phi_k$ is a Cauchy sequence in the finite dimensional space $\mathbb{C}^{(2K+1)^N}$, we get that for $k$ and $l$ sufficiently large and for any $\epsilon_2>0$, holds that
\begin{eqnarray}
\label{lem4}
\sum_{|n|\leq K}\delta_n|(\phi_k)_n-(\phi_l)_n|^p < \epsilon_2.
\end{eqnarray}
Inequality (\ref{lem2}) can be rewritten as
\begin{eqnarray}
\label{lem3}
||\phi_k-\phi_l||^{2p}_{\ell^2_{\delta}}\leq c\left\{\sum_{|n|\leq K}\delta_n|(\phi_k)_n-(\phi_l)_n|^p
+\sum_{|n|> K}\delta_n|(\phi_k)_n-(\phi_l)_n|^p\right\}.
\end{eqnarray}
Now from (\ref{lem5})-(\ref{lem3}), and appropriate choices of $\epsilon_1$ and $ \epsilon_2$,  we may derive that for sufficiently large $k$ and $l$,  $$||\phi_k-\phi_l||_{\ell^2_{\delta}}<\epsilon,\;\;\mbox{for any}\;\;\epsilon>0.$$ 
That is $\phi_k$ is a Cauchy sequence in $\ell^2_{\delta}$.\ \ $\diamond$
\vspace{0.2cm}

Let $\mathbf{A}:D(\mathbf{A})\subseteq X\rightarrow X$  a $\mathbb{C}$-linear, self-adjoint\ $\leq 0$ operator with dense domain $D(\mathbf{A})$ on the Hilbert space $X$, equipped with the scalar product $(\cdot ,\cdot)_{X}$. The space
$X_{\mathbf{A}}$ is the completion of $D(\mathbf{A})$ in the norm $||u||_{\mathbf{A}}^2=||u||^2_X-(\mathbf{A}u,u)_X$, for $u\in X_{\mathbf{A}}$, and we denote
by $X_{\mathbf{A}}^*$ its dual and by $\mathbf{A}^*$ the extension of $\mathbf{A}$ to the dual of $D(\mathbf{A})$, denoted by $D(\mathbf{A})^*$ (Friedrichs extension theory \cite{Davies1}, \cite[Vol. II/A]{zei85}).

Considering the operator $\mathbf{A}$ defined by (\ref{DiscLap}), we observe that for any $\phi\in\ell^2$
\begin{eqnarray}
\label{preA}
||\mathbf{A}\phi||_{\ell^2}^2\leq 4N||\phi||_{\ell^2}^2,
\end{eqnarray}
that is, $\mathbf{A}:\ell^2\rightarrow\ell^2$ is a continuous operator. Now we consider the discrete operator $\mathbf{L}^+:\ell^2\rightarrow\ell^2$  defined by
\begin{eqnarray}
\label{discder1}
(\mathbf{L}^+\psi)_{n\in\mathbb{Z}^N}&=&\left\{\psi_{(n_{1}+1,n_2,\ldots ,n_N)}-\psi_{(n_{1},n_2,\ldots ,n_N)}\right\}\nonumber\\
&+&\left\{\psi_{(n_{1},n_2+1,\ldots ,n_N)}-\psi_{(n_{1},n_2,\ldots ,n_N)}\right\}\nonumber\\
&\vdots&\nonumber\\
&+&\left\{\psi_{(n_{1},n_2,\ldots ,n_N+1)}-\psi_{(n_{1},n_2,\ldots ,n_N)}\right\},
\end{eqnarray}
and $\mathbf{L}^{-}:\ell^2\rightarrow\ell^2$ defined by
\begin{eqnarray}
\label{discder2}
(\mathbf{L}^-\psi)_{n\in\mathbb{Z}^N}&=&\left\{\psi_{(n_{1}-1,n_2,\ldots ,n_N)}-\psi_{(n_{1},n_2,\ldots ,n_N)}\right\}\nonumber\\
&+&\left\{\psi_{(n_{1},n_2-1,\ldots ,n_N)}-\psi_{(n_{1},n_2,\ldots ,n_N)}\right\}\nonumber\\
&\vdots&\nonumber\\
&+&\left\{\psi_{(n_{1},n_2,\ldots ,n_N-1)}-\psi_{(n_{1},n_2,\ldots ,n_N)}\right\}.
\end{eqnarray}
Setting
\begin{eqnarray}
\label{discder3}
(\mathbf{L}^+_{\nu}\psi)_{n\in\mathbb{Z}^N}=\psi_{(n_{1},n_2,\ldots , n_{\nu-1},n_{\nu}+1,n_{\nu+1},\ldots ,n_N)}-\psi_{(n_{1},n_2,\ldots ,n_N)},\\
\label{discder4}
(\mathbf{L}^-_{\nu}\psi)_{n\in\mathbb{Z}^N}=\psi_{(n_{1},n_2,\ldots , n_{\nu-1},n_{\nu}-1,n_{\nu+1},\ldots ,n_N)}-\psi_{(n_{1},n_2,\ldots ,n_N)},
\end{eqnarray}
we observe that the operator $\mathbf{A}$ satisfies the relations
\begin{eqnarray}
\label{diffop2}
(-\mathbf{A}\psi_1,\psi_2)_{\ell^2}&=&\sum_{\nu=1}^N(\mathbf{L}_\nu^+\psi_1,\mathbf{L}_\nu^+\psi_2)_{\ell^2},\;\;\mbox{for all}\;\;\psi_1,\psi_2\in\ell^2,\\
\label{diffop3}
(\mathbf{L}_\nu^+\psi_1,\psi_2)_{\ell^2}&=&(\psi_1,\mathbf{L}_\nu^-\psi_2)_{\ell^2},\;\;\mbox{for all}\;\;\psi_1,\psi_2\in\ell^2.
\end{eqnarray}
From (\ref{diffop2}), it is clear that $\mathbf{A}:\ell^2\rightarrow\ell^2$ defines a self adjoint operator on $\ell^2$,  and $\mathbf{A}\leq 0$. The graph-norm 
\begin{eqnarray*}
||\phi||_{D(\mathbf{A})}^2=||\mathbf{A}\phi||_{\ell^2}^2+||\phi||_{\ell^2}^2,
\end{eqnarray*}
is an equivalent with that of $\ell^2$, since
\begin{eqnarray*}
||\phi||_{\ell^2}^2\leq ||\phi||^2_{D(\mathbf{A})}\leq (4N+1)||\phi||_{\ell^2}^2.
\end{eqnarray*}
In our case, it appears that $X_{\mathbf{A}}=\ell^2$
equipped with the norm 
$$||\phi||_{\mathbf{A}}^2=||\phi||_X^2-(\mathbf{A}\phi,\phi)_X=\sum_{\nu=1}^{N}||\mathbf{L}^+_{\nu}\phi||_{\ell_2}^2+ ||\phi||^2_{\ell^2},$$ 
for $\phi\in\ell^2$, and is an equivalent norm with the usual one of $\ell^2$. Moreover,
$D(\mathbf{A})=X=\ell^2=D(\mathbf{A})^*$. Obviously $\mathbf{A}^*=\mathbf{A}$ and the operator
$i\mathbf{A}:\ell^2\rightarrow \ell^2$ defined by $(i\mathbf{A})\phi=i\mathbf{A}\phi$ for $\phi\in
\ell^2$, is $\mathbb{C}$-linear and skew-adjoint and $i\mathbf{A}$
generates a group $(\mathcal{T}(t))_{t\in\mathbb{R}}$, of
isometries on $\ell^2$ (see \cite{cazS}). 
The  analysis of the operator $\mathbf{A}$ is useful if one would like to consider the DNLS equation (\ref{DNLSh}) as an abstract evolution equation \cite{AN}, and holds for other discrete operators
which are not necessary discretizations of the Laplacian (for example as those of \cite{SZ2}).
%%%%%%%%%%%%%%%%%%%%%%%%%%%%%%%%%%%%%%%%%%%%%%%%%%%%%%%%%%%%%%%
%%%%%%%%%%%%%%%%%%%%%%%%%%%%%%%%%%%%%%%%%%%%%%%%%%%%%%%%%%%%%%%
%%%%%%%%%%%%%%%%%%%%%%%%%%%%%%%%%%%%%%%%%%%%%%%%%%%%%%%%%%%%%%%
\subsection{Existence of non trivial breather solutions in the case of decaying anharmonic parameter} 
We shall seek for nontrivial breather solutions as critical points of the functional
\begin{eqnarray}
\label{Enegfun}
\mathbf{E}(\phi )=\frac{1}{2}\sum_{\nu=1}^{N}||\mathbf{L}^+_{\nu}\phi||_{\ell_2}^2+\frac{\omega^2}{2}\sum_{n\in\mathbb{Z}^N}|\phi_n|^2-\frac{1}{2\sigma +2}\sum_{n\in\mathbb{Z}^N}\gamma_n|\phi_n|^{2\sigma +2}.
\end{eqnarray}
To establish differentiability of the functional
$\mathbf{E}:\ell^2\rightarrow\mathbb{R}$, we shall use the
following discrete version of the dominated convergence Theorem,
provided by \cite{Bates2}.
\begin{theorem}
\label{dc}
Let $\{\psi_{i,k}\}$ be a double sequence of summable functions, $$\sum_{i\in\mathbb{Z}^N}|\psi_{i,k}|<\infty,$$ 
and $\lim_{k\rightarrow\infty}\psi_{i,k}=\psi_{i}$, for all
$i\in\mathbb{Z}^N$. If there exists a summable sequence $\{g_{i}\}$ such that $|\psi_{i,k}|\leq g_{i}$ for all $i,k$'s, we have that
$\lim_{k\rightarrow\infty}\sum_{i\in\mathbb{Z}^N}\psi_{i,k}=\sum_{i\in\mathbb{Z}^N}\psi_{i}$.
\end{theorem}
We then have the following Lemma.
\begin{lemma}
\label{derivative}
Let $(\phi_n)_{n\in\mathbb{Z}^N}=\phi\in\ell^{2\sigma+2}$ for some $0<\sigma <\infty$. Moreover we assume that  $\gamma_n\in\ell^{\rho}$, $\rho=\frac{q-1}{q-2}$ for some $q>2$. Then the functional
$$\mathbf{F}(\phi)=\sum_{n\in\mathbb{Z}^N}\gamma_n|\phi_n|^{2\sigma +2},$$
is a $\mathrm{C}^{1}(\ell^{2\sigma +2},\mathbb{R})$ functional and
\begin{eqnarray}
\label{gatdev}
<\mathbf{F}'(\phi),\psi>=(2\sigma +2)\mathrm{Re}\sum_{n\in\mathbb{Z}^N}\gamma_n|\phi_n|^{2\sigma}\phi_n\overline{\psi_n},\;\;\psi=(\psi_n)_{n\in\mathbb{Z}^N}\in\ell^{2\sigma +2}.
\end{eqnarray}
\end{lemma}
{\bf Proof:}\ \ We assume that $\phi,\,\psi\in\ell^{2\sigma +2}$.
Then for any $n\in\mathbb{Z}^N$, $0<s<1$, we get 
\begin{eqnarray}
\label{mv}
&&\frac{\mathbf{F}(\phi +s\psi)-\mathbf{F}(\psi)}{s}=\frac{1}{s}\mathrm{Re}\sum_{n\in\mathbb{Z}^N}\gamma_n\int_{0}^{1}\frac{d}{d\theta}|\phi_n +
\theta s\psi_n|^{2\sigma +2}d\theta\\
&&\;\;\;\;\;\;\;\;\;\;\;\;\;\;\;\;\;\;\;\;\;\;=(2\sigma +2)\mathrm{Re}\sum_{n\in\mathbb{Z}^N}\gamma_n\int_{0}^{1}|\phi_n+s\theta\psi_n|^{2\sigma}
(\phi_n+s\theta\psi_n)\overline{\psi_n} d\theta.\nonumber
\end{eqnarray}
Since $\gamma_n$ is in $\ell^{\rho}$ it follows from (\ref{lp1}) that  
\begin{eqnarray}
\label{boundV}
\mathrm{sup}_{n\in\mathbb{Z}^N}|\gamma_n|=M<\infty.
\end{eqnarray}
On the other hand we have the  inequality
\begin{eqnarray}
\label{mv1}
&&\sum_{n\in\mathbb{Z}^N}|\phi_n+\theta s\psi_n|^{2\sigma +1}|\psi_n|
\leq
\sum_{n\in\mathbb{Z}^N}\left(|\phi_n|+|\psi_n|\right)^{2\sigma +1}|\psi_n|\;\;\;\;\;\;\;\nonumber\\
\;\;\;\;\;\;\;\;\;\;\;\;\;\;\;&&\leq\left(\sum_{n\in\mathbb{Z}^N}(|\phi_n|+|\psi_n|)^{2\sigma +2}\right)^{\frac{2\sigma +1}{2\sigma +2}}
\left(\sum_{n\in\mathbb{Z}^N}|\psi_n|^{2\sigma +2}\right)^{\frac{1}{2\sigma +2}}.
\end{eqnarray}
Now by using (\ref{boundV}) and inserting (\ref{mv1}) into (\ref{mv}), we see that Lemma \ref{dc} is applicable: Letting $s\rightarrow 0$, we get the existence of the Gateaux derivative (\ref{gatdev}) of the functional $\mathbf{F}:\ell^{2\sigma +2}\rightarrow\mathbb{R}$.

We show next that the functional $\mathbf{F}':\ell^{2\sigma +2}\rightarrow\ell^{\frac{2\sigma +2}{2\sigma +1}}$ is
continuous. For $\phi\in\ell^{2\sigma +2}$, we set $(F_1(\phi))_{n\in\mathbb{Z}^N}=|\phi_n|^{2\sigma}\phi_n$. 

Let us note that for any $F\in \mathrm{C}(\mathbb{C},\mathbb{C})$ which takes the form
$F(z)=g(|z|^2)z$, with  $g$ real and sufficiently smooth, holds 
\begin{eqnarray}
\label{GL}
F(\phi_1)-F(\phi_2)=\int_{0}^{1}\left\{(\phi_1-\phi_2)(g(r)+rg'(r))+(\overline{\phi}_1-\overline{\phi}_2)\Phi^2 g'(r)\right\}d\theta,
\end{eqnarray}
for any $\phi_1,\;\phi_2\in \mathbb{C}$,where $\Phi=\theta \phi_1+(1-\theta)\phi_2$, $\theta\in (0,1)$ and $r=|\Phi|^2$ (see \cite[pg. 202]{GiVel96}).
Applying (\ref{GL}) for the case of $F_1$ $(g(r)= r^{\sigma})$ one obtains that
\begin{eqnarray*}
F_1(\phi_1)-F_1(\phi_2)=\int_0^1[(\sigma
+1)(\phi_1-\phi_2)|\Phi|^{2\sigma}
+\sigma(\overline{\phi}_1-\overline{\phi}_2)\Phi^2|\Phi|^{2\sigma
-2}]d\theta,
\end{eqnarray*}
which implies the inequality
\begin{eqnarray}
\label{GL2}
|F_1(\phi_1)-F_1(\phi_2)|\leq (2\sigma +1)(|\phi_1|+|\phi_2|)^{2\sigma}|\phi_1-\phi_2|.
\end{eqnarray}
We consider a sequence $\phi_m\in\ell^{2\sigma +2}$ such that $\phi_m\rightarrow \phi$ in $\ell^{2\sigma +2}$. Using (\ref{boundV}), we get the inequality
\begin{eqnarray}
\label{hoin}
\left|\left<\mathbf{F}'(\phi_m)-\mathbf{F}'(\phi),\,\psi\right>\right|&\leq& c(M)||F_1(\phi_m)-F_1(\phi)||_{\ell^{q}}||\psi||_{\ell^p},\\
&&q=\frac{2\sigma +2}{2\sigma +1},\;\;p=2\sigma +2.\nonumber
\end{eqnarray}
We denote by $(\phi_m)_n$ the $n$-th coordinate of the sequence $\phi_m\in\ell^2$. By setting  $\Phi_n=(|(\phi_m)_n|+|\phi_n|)^{2\sigma}$, we get from (\ref{GL2}), that for some constant $c>0$
\begin{eqnarray*}
&&\sum_{n\in\mathbb{Z}^N}|F_1((\phi_m)_n)-F_1(\phi_n)|^{q}\leq c\sum_{n\in\mathbb{Z}^N}(\Phi_n)^q|(\phi_m)_n-\phi_n|^{q}\nonumber\\
&&\leq c\left(\sum_{n\in\mathbb{Z}^N}|(\phi_m)_n-\phi_n|^{2\sigma +2}\right)^{\frac{1}{2\sigma +1}}
\left(\sum_{n\in\mathbb{Z}^N}(\Phi_n)^{\frac{\sigma +1}{\sigma}}\right)^{\frac{2\sigma}{2\sigma +1}}\rightarrow 0,
\end{eqnarray*}
as $m\rightarrow\infty$.\ \ \ $\diamond$

By using (\ref{diffop2}), we may easily get that the rest of the terms of the functional $\mathbf{E}$ given by (\ref{Enegfun}), define $\mathrm{C}^1(\ell^2,\mathbb{R})$ functionals. Since Lemma \ref{derivative} holds for any $\phi\in\ell^2$ (by (\ref{lp1})), we finally obtain that the functional $\mathbf{E}$ is $\mathrm{C}^1(\ell^2,\mathbb{R})$. Moreover, by using the analysis of Section 1 for the self-adjoint operator $\mathbf{A}:\ell^2\rightarrow\ell^2$, it appears that  any solution of 
(\ref{swe}), satisfies the formula
\begin{eqnarray*}
(-\mathbf{A}\phi,\psi)_{\ell^2}+\omega(\phi,\psi)_{\ell^2}=(\gamma_nF_1(\phi),\psi)_{\ell^2},\;\;\mbox{for all}\;\;\psi\in\ell^2,
\end{eqnarray*}
and vice versa. Equivalently, due to the differentiability of the functional $\mathbf{E}$, any solution of (\ref{swe}) is a critical point of $\mathbf{E}$. For convenience, we recall \cite[Definition 4.1, pg. 130]{CJ}
(PS-condition) and \cite[Theorem 6.1, pg. 140]{CJ} or
\cite[Theorem 6.1, pg. 109]{struwe} (Mountain Pass Theorem of Ambrosseti-Rabinowitz \cite{Amb}). 
\begin{definition}
\label{condc} Let $X$ be a Banach space and $\mathbf{E}:X\rightarrow\mathbb{R}$ be $\mathrm{C}^1$. We say that
$\mathbf{E}$ satisfies condition $(PS)$ if, for any sequence $\{\phi_n\}\in X$ such that $|\mathbf{E}(\phi_n)|$ is bounded and $\mathbf{E}'(\phi_n)\rightarrow 0$ as $n\rightarrow\infty$,
there exists a convergent subsequence. If this condition is only satisfied in the region where $\mathbf{E}\geq\alpha >0$ (resp $\mathbf{E}\leq -\alpha <0$) for all $\alpha >0$, we say $\mathbf{E}$ satisfies condition $(PS^+)$ (resp. $(PS^-)$).
\end{definition}
\begin{theorem}
\label{mpass}
Let $\mathbf{E}:X\rightarrow\mathbb{R}$ be $C^1$ and satisfy (a) $\mathbf{E}(0)=0$, (b) $\exists\rho >0$, $\alpha >0:\;||\phi||_X=\rho$ implies $\mathbf{E}(\phi)\geq\alpha$, (c) $\exists \phi_1\in X :\;||\phi_1||_X\geq\rho$ and
$\mathbf{E}(\phi_1)<\alpha$. Define $$\Gamma=\left\{\gamma\in \mathrm{C}^0([0,1],X):\;\gamma (0)=0,\;\;\gamma (1)=\phi_1\right\}.$$
Let $F_{\gamma}=\{\gamma(t)\in X:\;0\leq t\leq 1\}$ and $\mathcal{L}=\{F_\gamma :\;\gamma\in \Gamma\}$. If $\mathbf{E}$ satisfies condition $(PS)$, then $$\beta:=\inf_{F_{\gamma}\in \mathcal{L}}\sup\{\mathbf{E}(v):v\in F_{\gamma}\}\geq\alpha$$ is a critical point of the functional $\mathbf{E}$.
\end{theorem}
For fixed $\omega>0$, we shall consider a norm in $\ell^2$ defined by
\begin{eqnarray}
\label{moup1}
||\phi||_{\ell^2_{\omega}}^2=\sum_{\nu=1}^{\nu=N}||\mathbf{L}^+_{\nu}\phi||_{\ell_2}^2+\omega ||\phi||^2_{\ell^2},\;\;\phi\in\ell^2.
\end{eqnarray}
The norm (\ref{moup1}) is an equivalent norm with the usual one of $\ell^2$, since
\begin{eqnarray}
\label{moup2}
\omega ||\phi||^2_{\ell^2}\leq ||\phi||_{\ell^2_\omega}^2\leq (2N+\omega)||\phi||^2_{\ell^2}.
\end{eqnarray} 
We first check the behavior of the functional $\mathbf{E}$. Using (\ref{moup2}), we observe that
\begin{eqnarray}
\label{moup3}
|\mathbf{F}(\phi)|&\leq& M\sum_{n\in\mathbb{Z}^N}|\phi_n|^{2\sigma+2}
\leq M||\phi||_{\ell^2}^{2\sigma+2}\nonumber\\
&\leq&\frac{M}{\omega^{\sigma+1}}||\phi||^{2\sigma+2}_{\ell^2_{\omega}}.
\end{eqnarray}
Now setting $M_0=M/\omega^{\sigma+1}$
we observe that
\begin{eqnarray}
\label{moup4}
\mathbf{E}(\phi)&=&\frac{1}{2}||\phi||^2_{\ell^2_{\omega}}-\frac{1}{2\sigma+2}\mathbf{F}(\phi)\nonumber\\
&&\geq \frac{1}{2}||\phi||^2_{\ell^2_{\omega}}-\frac{M_0}{2\sigma+2}||\phi||^{2\sigma+2}_{\ell^2_\omega}.
\end{eqnarray}
Now we select some $\phi\in\ell^2$ such that $||\phi||_{\ell^2_{\omega}}=R>0$. Then, if 
\begin{eqnarray}
\label{disp1}
0<R < \left(\frac{\sigma +1}{M_0}\right)^{\frac{1}{2\sigma}}=\left(\frac{(\sigma +1)\omega^{\sigma+1}}{M}\right)^{\frac{1}{2\sigma}}:=E_{\ell^2_{\omega}}^*(\sigma,\omega,M), 
\end{eqnarray}
it follows from (\ref{moup4}) that
\begin{eqnarray}
\label{moup5}
\mathbf{E}(\phi) \geq \alpha>0,\;\;\alpha=R^2\left(\frac{1}{2}-\frac{M_0}{2\sigma +2}R^{2\sigma}\right).
\nonumber
\end{eqnarray}
We assume that $\gamma_n>0$ for all $n\in\mathbf{S}_+\subseteq\mathbb{Z}^N$. We shall consider next, some $\psi\in\ell^2$ such that $||\psi||_{\ell^2_{\omega}}=1$ and
\begin{eqnarray*}
\{\psi_n\}_{n\in\mathbb{Z}}=\{\psi_n\}_{n\in\mathbf{S}_+}+\{\psi_n\}_{n\in(\mathbb{Z}^N\setminus\mathbf{S}_+)},\;\;\mbox{where}\;\;
\left\{
\begin{array}{rlr}
&\{\psi_n\}_{n\in\mathbf{S}_+}&>0,\nonumber \\
&\{\psi_n\}_{n\in(\mathbb{Z}^N\setminus\mathbf{S}_+)}&=0.
\end{array}
\right.
\end{eqnarray*}
For some $t>0$ we considet the element $\chi=t\psi\in\ell^2$. We have that
\begin{eqnarray}
\label{moup6}
\mathbf{E}(\chi)=\frac{t^2}{2}-\frac{1}{2\sigma +2}t^{2\sigma +2}\sum_{n\in\mathbf{S}_+}\gamma_n|\psi_n|^{2\sigma+2}.
\end{eqnarray}
Now letting $t\rightarrow +\infty$ we get that $\mathbf{E}(t\psi)\rightarrow -\infty$. 

For fixed $\phi\neq 0$ and choosing $t$ sufficiently large, we may set $\phi_1=t\phi$ to satisfy the second condition of Theorem \ref{mpass}. To conclude with the existence of a non-trivial breather solution, it remains to show that the functional $\mathbf{E}$ satisfies Lemma \ref{condc}.

To this end, we consider a sequence $\phi_m$ of
$\ell^2$ be such that $|\mathbf{E}(\phi_m)|<M'$ for some $M'>0$ and
$\mathbf{E}'(\phi_m)\rightarrow 0$ as $m\rightarrow\infty$. By using
(\ref{Enegfun}) and Lemma \ref{derivative}, we observe that for $m$ sufficiently large
\begin{eqnarray}
\label{boundP.S}
M'\geq \mathbf{E}(\phi_m)-\frac{1}{2\sigma +2}\left<\mathbf{E}'(\phi_m),\phi_m\right>=
\left(\frac{1}{2}-\frac{1}{2\sigma +2}\right)||\phi_m||^2_{\ell^2_{\omega}}.
\end{eqnarray}
Therefore the sequence $\phi_m$ is bounded. Thus, we may extract a subsequence, still denoted by $\phi_m$, such that
\begin{eqnarray}
\label{weakcon}
\phi_m\rightharpoonup \phi\;\;\mbox{in}\;\;\ell^2,\;\;\mbox{as}\;\;m\rightarrow\infty.
\end{eqnarray}
For this subsequence it follows once again from (\ref{Enegfun}) and Lemma \ref{derivative} that
\begin{eqnarray}
\label{moup8}
||\phi_m-\phi||_{\ell^2_{\omega}}^2&=&\left<\mathbf{E}'(\phi_m)-\mathbf{E}'(\phi),\phi_m-\phi\right>\nonumber\\
&&+\sum_{n\in\mathbb{Z}^N}\gamma_n[|(\phi_m)_n|^{2\sigma}(\phi_m)_n-|\phi_n|^{2\sigma}\phi_n]((\phi_m)_n-\phi_n)).\;\;\;
\end{eqnarray}
Another assumption on the sequence $\gamma_n$ is that the sequence $|\gamma_n|=(\delta_n)_{n\in\mathbb{Z}^N}$ satisfies the assumptions of Lemma \ref{compactness}. We consider the associated Hilbert space $\ell^2_{\delta}$. Now by using the inequality (\ref{GL2}), we get for the second term of right hand side of (\ref{moup8}), the estimate
\begin{eqnarray}
\label{moup9}
&&\sum_{n\in\mathbb{Z}^N}\gamma_n[|(\phi_m)_n|^{2\sigma}(\phi_m)_n-|\phi_n|^{2\sigma}\phi_n]((\phi_m)_n-\phi_n))\;\;\;\;\;\nonumber\\
&&\;\;\;\;\leq c\sum_{n\in\mathbb{Z}^N}\Phi_n|\gamma_n|\,\,|(\phi_m)_n-\phi_n|^2\nonumber\\
&&\;\;\;\;\leq c\sup_{n\in\mathbb{Z}^N}\Phi_n\sum_{n\in\mathbb{Z}^N}|\gamma_n|\,|(\phi_m)_n-\phi_n|^2
=c_2||\phi_m-\phi||_{\ell^2_\delta}^2,
\end{eqnarray}
where $c_2=c\sup_{n\in\mathbb{Z}^N}\Phi_n$. 
Obviously, $\phi_m$ is bounded in $\ell^2_{\delta}$ and by Lemma \ref{compactness} it follows that
\begin{eqnarray}
\label{weakcon2}
\phi_m\rightarrow \phi\;\;in\;\;\ell^2_{\delta},\;\;\mbox{as}\;\;m\rightarrow\infty.
\end{eqnarray}
Combining (\ref{moup8}), (\ref{moup9}) and (\ref{weakcon2}), we obtain that $$||\phi_m-\phi||_{\ell^2_{\omega}}\rightarrow 0,\;\;\mbox{as}\;\;m\rightarrow\infty.$$ Hence $\phi_m$ has a (strongly) convergent subsequence.  The assumptions of Theorem \ref{mpass} are satisfied, and we may summarize in the following
\begin{theorem}
\label{COMP}
Assume that the site-dependent anharmonic parameter $\gamma_n>0$ in some $\mathbf{S}_+\subseteq\mathbb{Z}^N$. Moreover, we assume that $|\gamma_n|=\delta_n\in\ell^{\rho}$, $\rho=\frac{q-1}{q-2}$ for some positive integer $q>2$. Then for any $\omega>0$, there exists a nontrivial breather solution $\psi_n(t)=\phi_n\exp (i\omega t)$ of the DNLS equation (\ref{DNLSh}).
\end{theorem}
\vspace{0.2cm}

We remark here that the assumptions on the sequence of anharmonic parameters $\gamma_n$, are crucial for the derivation of the strong convergence of the subsequence $\phi_m$. If $\gamma_n$ is constant for all $n\in\mathbb{Z}^N$, then due to the lack of the Schur property for the space $\ell^{2}$ (in contrast with the space $\ell^1$ which posses this property-weak convergence coincides with strong convergence), we may not conclude the strong convergence of the subsequence, from its weak convergence. Of course the strong convergence, is valid in the case of a finite lattice: In this case, the problem is formulated in finite dimensional spaces where weak is equivalent to strong convergence \cite{AN}. 

Inequality (\ref{disp1}) could have some physical interpretation with respect to the  nontrivial breather solutions of frequency $\omega>0$, if one considers (\ref{disp1}) as a possible local upper bound for the ``energy'' quantity defined by (\ref{moup1}). It contains information on the type of nonlinearity and the sequence of anharmonic parameters, through its dependence on the nonlinearity exponent $\sigma$ and $M$. Such type of relations seem to be reasonable, as the next result concerning nonexistence of nontrivial breather solutions shows. The restriction on the energy of the excitations for nonexistence, combined with the upper bound (\ref{disp1}) above, could  provide us with some indicative information, on the behavior of energy quantities, associated with the nontrivial breather solution.
%%%%%%%%%%%%%%%%%%%%%%%%%%%%%%%%%%%%%%%%%%%%%5

For the shake of completeness, we recall 
\cite[Theorem 18.E, pg. 68]{zei85}
(Theorem of Lax and Milgram). This theorem will be used to establish existence of solutions for an auxiliary infinite linear system of algebraic equations related to (\ref{swe}). 
\begin{theorem}
\label{LMth} 
Let $X$ be a Hilbert space and $\mathbf{A}:X\rightarrow X$ be a linear continuous operator. Suppose that there exists  $c^*>0$ such that
\begin{eqnarray}
\label{strongmonot}
\mathrm{Re}(\mathbf{A}u,u)_X\geq c^*||u||^2_X,\;\;\mbox{for all}\;\;u\in X.
\end{eqnarray}
Then for given $f\in X$, the operator equation $\mathbf{A}u=f,\;\;u\in X$, has a unique solution
\end{theorem}
The non existence result can be stated as follows
\begin{theorem}
\label{notri}
There exist no nontrivial breather solution  of energy less than 
\begin{eqnarray}
\label{disp2}
E_{\mathrm{min}}(\omega,\sigma,M):=\frac{1}{2}\left(\frac{\omega}{M(2\sigma +1)}\right)^{1/2\sigma}.
\end{eqnarray}
\end{theorem}
{\bf Proof:}\ \  For some $\omega> 0$, we consider the operator $\mathbf{A}_{\omega}:\ell^2\rightarrow\ell^2$, defined by
\begin{eqnarray}
\label{strongop1}
(\mathbf{A}_{\omega}\phi)_{n\in\mathbb{Z}^N}&=&(\mathbf{A}\phi)_{n\in\mathbb{Z}^N}+\omega\phi_n.
\end{eqnarray}
It is linear and continuous and satisfies assumption (\ref{strongmonot}) of Theorem \ref{LMth}: Using (\ref{diffop2}), we get that
\begin{eqnarray}
\label{check}
(\mathbf{A}_{\omega}\phi,\phi)_{\ell^2}=\sum_{\nu=1}^N|\mathbf{L}^+_{\nu}\phi||^2_{\ell^2}+\omega ||\phi||^2\geq \omega ||\phi||^2_{\ell^2}\;\;\mbox{for all}\;\;\phi\in\ell^2.
\end{eqnarray}
For given $z\in\ell^2$, we consider the linear operator equation
\begin{eqnarray}
\label{linear}
(\mathbf{A}_{\omega}\phi)_{n\in\mathbb{Z}^N}=\gamma_n|z_n|^{2\sigma}z_n.
\end{eqnarray}
For the map 
\begin{eqnarray}
\label{nolimap}
(\mathbf{T}(z))_{n\in\mathbb{Z}^N}=\gamma_n|z_n|^{2\sigma}z_n,
\end{eqnarray}
we observe that
\begin{eqnarray*}
||\mathbf{T}(z)||^2_{\ell^2}\leq M^2\sum_{n\in\mathbb{Z}^N}|z_n|^{4\sigma +2}
\leq M^2||z||_{\ell^2}^{4\sigma +2}.
\end{eqnarray*}
Hence, the assumptions of Theorem \ref{LMth} are satisfied, and (\ref{linear}) has a unique solution $\phi\in\ell^2$. For some $R>0$, we consider the closed ball of $\ell^2$, $B_R:=\{z\in\ell^2\;:||z||_{\ell^2}\leq R\}$, and we define the map
$\mathcal{P}:\ell^2\rightarrow\ell^2$, by $\mathcal{P}(z):=\phi$ where $\phi$ is the unique solution of the operator equation (\ref{linear}). Clearly the map $\mathcal{P}$ is well defined. 

Let $\zeta$, $\xi\in B_R$ such that $\phi=\mathcal{P}(\zeta)$, $\psi=\mathcal{P}(\xi)$. The difference $\chi:=\phi-\psi$ satisfies the equation
\begin{eqnarray}
\label{claim2} 
(\mathbf{A}_{\omega}\chi)_{n\in\mathbb{Z}^N}=(\mathbf{T}(z))_{n\in\mathbb{Z}^N}-(\mathbf{T}(\xi))_{n\in\mathbb{Z}^N}.
\end{eqnarray}
The map $\mathbf{T}:\ell^2\rightarrow\ell^2$ is locally Lipschitz, since we may use (\ref{GL2}) once again, to get
\begin{eqnarray}
\label{claim3}
||\mathbf{T}(\zeta)-\mathbf{T}(\xi)||_{\ell^2}^2&\leq& (2\sigma+1)^2M^2\sum_{n\in\mathbb{Z}^N}(|\zeta_n|+|\xi_n|)^{2\sigma})^2|\zeta_n-\xi_n|^2\nonumber\\
&\leq&(2\sigma+1)^2M^2[\sup_{n\in\mathbb{Z}^N}(|\zeta_n|+|\xi_n|)^{2\sigma})]^2\sum_{n\in\mathbb{Z}^N}|\zeta_n-\xi_n|^2\nonumber\\
&\leq& M_1^2R^{4\sigma}||\zeta-\xi||^2_{\ell^2},
\end{eqnarray}
whith $M_1=2^{2\sigma}M(2\sigma+1)$. Taking now the scalar product of (\ref{claim2}) with $\chi$ in $\ell^2$ and using (\ref{claim3}), we have
\begin{eqnarray}
\label{cmap1a}
\sum_{\nu=1}^N||\mathbf{L}_{\nu}^+\chi||^2_{\ell^2}+\omega ||\chi||^2_{\ell^2}&\leq& ||\mathbf{T}(\zeta)-\mathbf{T}(\xi)||_{\ell^2}||\chi||_{\ell^2}\nonumber\\
&\leq&M_1R^{2\sigma}||\zeta-\xi||_{\ell^2}||\chi||_{\ell^2}\nonumber\\
&\leq&\frac{\omega}{2}||\chi||_{\ell^2}^2+\frac{1}{2\omega}M^2_1R^{4\sigma}||z-\xi||_{\ell^2}^2.
\end{eqnarray}
From (\ref{cmap1a}), we obtain the inequality
\begin{eqnarray}
\label{claim4}
||\chi||_{\ell^2}^2=||\mathcal{P}(z)-\mathcal{P}(\xi)||_{\ell^2}^2
\leq \frac{1}{\omega^2}M^2_1R^{4\sigma}||z-\xi||^2_{\ell^2}.
\end{eqnarray}
Since $\mathcal{P}(0)=0$, from inequality  (\ref{claim4}), we derive that for $R< E_{\mathrm{min}}$,  the map  $\mathcal{P}:B_R\rightarrow B_R$ and is a contraction.   Therefore  $\mathcal{P}$, satisfies the assumptions of Banach Fixed Point Theorem and has a unique fixed point, the trivial one. Hence, for $R<E_{\mathrm{min}}$ the only breather solution is the trivial. \ \ $\diamond$   
\vspace{0.2cm}
\newline

If the energy of the excitation
is less that $E_{\mathrm{min}}$ the lattice may not support a standing wave of frequency $\omega$.  This time, relation (\ref{disp2}) could be seen as some kind of dispersion relation of frequency vs energy for the nonexsistence of breather solutions of the DNLS equation (\ref{DNLSh}). The dependence  $E_{\ell^2_{\omega}}^*$ and $E_{min}$ on $\omega,\sigma, M$ as it appears from inequalities (\ref{disp1}), (\ref{disp2}), could be a point of departure for  investigations on the relation of the energy quantity defined by (\ref{moup1}) and the $\ell^2$-norm of the nontrivial breather solution (the power), as well as on their behavior.  For example, the inequality
$E_{\mathrm{min}}<E_{\ell^2_{\omega}}^*$,
is satisfied if
\begin{eqnarray}
\label{disp3}
\left(\frac{1}{2^{2\sigma}(\sigma+1)(2\sigma+1)}\right)^{\frac{1}{\sigma}}<\omega.
\end{eqnarray}
In the case $\sigma=1$ (cubic nonlinearity) we get a lower bound $\omega>24^{-1}\sim 0.04166$, for the frequency of the nontrivial breather solution $\psi_n(t)=\phi_n\exp (i\omega t)$, satisfying
\begin{eqnarray}
\label{disp4}
||\phi||_{\ell^2_{\omega}}>E_{\mathrm{min}}.
\end{eqnarray}

Let us also remark that a similar nonexistence result as Theorem \ref{notri}, can be proved in the case (a) of an  infinite lattice with $\gamma=\mathrm{const}, \cite{AN}$ and (b) the case of finite lattice (assuming Dirichlet boundary conditions).

Numerical simulations, for testing restriction (\ref{disp2}) or (\ref{disp3})-(\ref{disp4}), could be of  interest. Further developments could consider  DNLS equations with site dependence on the coupling strength, or operators which are not necessarily discretizations of the Laplacian (for examples of such operators see \cite{SZ2}). \vspace{0.2cm} 

{\bf Acknowledgements}. I would like to thank Professors J. C. Eilbeck, and J. Cuevas, for their valuable discussions (especially for resolving the significance of relation (\ref{disp3})) and their interest, improving considerably the presentation of the final version of the manuscript, and my colleagues A. N. Yannacopoulos and H. Nistazakis for their suggestions. I would like also to thank the referee for his useful comments. This work was partially supported by the research project proposal ``Pythagoras I-Dynamics of Discrete and Continuous Systems and Applications''- National Technical University of Athens and University of the Aegean.
%%%%%%%%%%%%%%%%%%%%%%%%%%%%%%%%%%%%%%%%%%%%%%%
%%%%%%%%%%%%%%%%%%%%%%%%%%%%%%%%%%%%%%%%%%%%%%%

\end{document}